\documentclass{llncs}
\usepackage[utf8]{inputenc}
\usepackage{multirow}
\usepackage{booktabs} % For formal tables
\usepackage{hyperref}
\usepackage{graphicx}

\title{Teaching Programming and Design-by-Contract}
\author{Daniel de Carvalho\inst{1} \and Rasheed Hussain \and Adil Khan\inst{1} \and Mansur Khazeev\inst{1} \and JooYong Lee\inst{1} \and Sergey Masiagin \and Manuel Mazzara\inst{1} \and Ruslan Mustafin\inst{1} \and Alexandr Naumchev\inst{1} \and Victor Rivera\inst{1}}

\institute{Innopolis University\\
\email{\{d.carvalho, r.hussain, a.khan, m.khazeev, j.lee, s.masiagin, m.mazzara, a.naumchev, v.rivera\}@innopolis.ru}
}

\authorrunning{<abbreviated author list>}

\begin{document}
\maketitle
\begin{abstract}
This paper summarizes the experience of teaching an introductory course to programming by using a correctness by construction approach at Innopolis University, Russian Federation. In this paper we claim that division in beginner and advanced groups improves the learning outcomes, present the discussion and the data that support the claim.
% We discuss the data supporting the idea that a division in beginner and advanced groups improves the learning outcomes.
\end{abstract}

%----------------------------------------------------
\section{Introduction}  \label{Intro}
%----------------------------------------------------

Formal methods are still struggling to get a broad acceptance in industry world-wide, and Russia  is not an exception. Innopolis is a new IT city \cite{Kondratyev:13}, incorporating a technopark and a university, aiming at prioritizing the development of IT and software engineering in Tatarstan and in the Russian Federation. Innopolis University (IU) is pioneering several research and pedagogical projects and experiments with innovative teaching methods and curricula. One of the numerous peculiarities of this innovation has been the decision to teach formal methods and correctness by construction together with programming since the first year of the bachelor program. In particular the Eiffel programming language \cite{Meyer:1992} is used as a programming instrument and Design by Contract as a methodological and conceptual tool \cite{Meyer:1992b}.

This paper summarizes the experience accumlated by followinng this pedagogical approach. It also reports on the course structure and answers questions related to setup, programming language and chosen paradigm. The work is structured as following: Section \ref{Eiffel} motivates the choice of adopting Eiffel as first programming language to be studied. 
%Section \ref{Millennials} presents our perspective on the key features of millennials, i.e. the student target we are teaching right now. 
Section \ref{DbC} describes the structure of the course and the Design by Contract approach adopted. Section \ref{Results} reports empirical results on our teaching effort: a poll was presented to students and some data collected. Finally, the numerical data presented is then analyzed and commented in Section \ref{Discussion}.

%----------------------------------------------------
\section{Eiffel as first language} 
\label{Eiffel}
%----------------------------------------------------

The choice of Eiffel and Design by Contract as programming and methodological tools for first year bachelors has been long discussed inside the university. After four years we could not find any evidence suggesting the need for a change. Instead, we will consider some data supporting the idea that the course worked out succesfully, both in terms of content and organization. It is however worth motivating the decision more in detail.

Which programming language is better to start studying for the beginners? No single answer exists. In general, it is easier to answer to the opposite question: "What programming language is better not to start with?". Experience has shown that teaching a specific language from scratch in order to satisfy a specific and urgent needs may not bring individuals to develop into a skilled and versatile professionals. Professional experience has shown cases of individuals who improvised themselves as Visual Basic programmers from scratch, or moved from FORTRAN or COBOL to Java because of some local business need. Often the immediate emergency was patched, but the correct mindset and basic skills required by an experienced professional were not developed. Sometimes such an emergency is inevitable, but developing a quality curriculum for a top-level university requires more care and deeper analysis of what programming is and programmers need, with some initial and pedagogical detachment from raw business needs.

Worldwide, examples of good pedagogical approaches for programming are not missing. There are a few preliminary considerations to be done in order to follow these succesfull steps . First, what programming paradigm we want to use? There is a general tendency to prefer the Object Oriented Programming (OOP) Paradigm \cite{Wegner:1990} as starting point since it helps students developing abstraction skills and design method. This approach, however, is not without its critics: some believe that Object Orientation may deal too much (and too early) with design and interface aspects and not enough with algorithmic details and imperative flow structure. According to this view, procedural programming would be better to start, while Object Orientation should be introduced in advanced courses. Of course, this depends on how the course is organized and taught, but the concern is serious. The school of thought privileging OOP usually concentrates on languages like Java \cite{java:00} or C\# \cite{Hejlsberg:2010} in order to take into account business demand. The school of procedural programming sometime concentrate on purely academic languages like Pascal \cite{Jensen:1974}, with the benefit of simplicity, or widespread languages like C \cite{Kernighan:1988}, offering broader flexibility (and related complexity). There are other paradigms too, for example the Functional (Lisp \cite{Steele:1990}, ML \cite{Milner:1997}, Scala \cite{Odersky:2004} and Haskell \cite{Jones:2003}), which has attracted renewed interest in recent years, and Logic (Prolog \cite{Bratko:1986}). 

The second general observation is that any computer scientist or software engineer will learn a number of programming languages over the course of his career. In particular, everyone will learn one or more of the dominant languages such as, today, Java, C\#, C, C++ or Python. So the choice of the first programming language is not exclusive of others; rather, it is a preparation for others, and should emphasize development of the skills needed to learn programming. (In fact, an increasing number of students have done some experience with Java or other languages before they even join the university program.) 

When desiging an introduction to programming course is also important to reflect on how much emphasis (if any) shoud be put on formal reasoning, software quality and correctness by construction. The University is the ideal time of life for learning new concepts and, at the same time, build the foundations of one's knowledge and mindset. Establishing a broad and deep basis is also the best way to make sure that students not only receive sufficient initial training to obtain a first job, but acquire the extensive long-term intellectual skills to pursue a successive career over several decades: the technologies will change, particularly in such a quickly evolving field as Information Technology, but the principles acquired during university study, if thought without dogma and with an open mind, will remain useful.

As a result, a broad school of thought supports the idea that the introductory programming course and the first programming language should emphasize Computer Science foundations and formal reasoning in order to strengthen a mindset leading to development of quality software. Eiffel and Design-by-Contract are just one possible technological and methodological solutions to implement such philosophy, and it is the one collectively chosen for our \textit{Introduction to Programming} in order to provide the adequate mindset to future professionals. This path is not free of controversy. The experience inherited from ETH Zurich \footnote{\url{https://www.ethz.ch/en.html}} is positive \cite{Pedroni:2006}, and the course was well received by students. We aim at repeating the success in different contexts, though an adaptation phase is necessary and benefits of the approach may not appear as immediate. 

% Overall, it is indeed easier to answer the opposite question, i.e. what is better not to do. Among the several possible paths, we have chosen one. However, as described above, worldwide several other equally valid decisions have been taken, plenty of courses delivered and resulted beneficial to beginners.

% There is no single correct answer to this question, no common agreement on this and debate in some cases is vibrant. There is no global consent between programmers, teachers, academics or business people. The reason is simple: the question is incorrectly formulated. For comparison, imagine that the question is not about programming languages but about human languages: what is the best foreign language to learn? Clearly, there is no single correct answer. The answer depends on such factors as the students' mother tongue, his cultural background, his objectives and motivation (professional or leisure?) and others. 

%The same goes for programming languages: 

%----------------------------------------------------
\section{Course Structure and approach} \label{DbC}
%----------------------------------------------------
In this section we will discuss how the course \textit{Introduction to Programming I} is structured at Innopolis.

The \textit{Introduction to Programming I} course (was called \textit{Object Oriented Programming} the first year it was delivered) is a 6 ETCS course delivered to first year bachelor students at IU over 15 weeks with 2 academic hours of frontal lectures and 4 hours of laboratory exercises every week. 
%There is also an \textit{Introduction to Programming II} course, but the nature and the approach as well as the lecturer of this course changed over the years. Instead \textit{OOP/Introduction to Programming I} kept the same structure. 
%Vic: at least for the revision part. We can add this if the paper is accepted
%, although the first year the lecturer was Manuel Mazzara, and the second and third year Victor Rivera. 
The team is composed by a Principal Instructor (PI) in charge of delivering lectures (PI has changed twice in the past years -- although they both work in the same team) and Teaching Assistants (TA) in charge of delivering laboratory exercises. PIs are formal method experts, and TAs are researchers or PhD students in the area. The foundational ideas on which the course is based are:

\begin{itemize}
\item The foundation for programming lies on mathematical and logical bases
\item Identifying and fixing bugs early is cost effective, hence the emphasis on correctness by construction (in synergy with testing)
\item Explain and delivering these points to the students so that this knowledge is passed to their future job environments
\end{itemize}

%All the team is very much supportive of the idea that the foundations for programming lies on the mathematical and logic bases, that identifying and fixing bugs early is cost effective, and that this message should be passed to students, and therefore brought into their job environments once they graduate. 

Frontal lectures are given in English to all students and there is no differentiation between the level of English proficiency of students. Lab sessions, on the other hand, are split into 4 categories: by the level of experience in programming, \textit{Beginners} or \textit{Advanced}, and by the natural language, \textit{Russian} (native language for the majority of students) and \textit{English}. Students had the ability to select which level they belong to based on their perception. The decision is done at the beginning of the course and students are not allowed to change once they have decided. 
%It is noticed that some students with less experience selected \textit{Advanced} groups for them to learn faster from more experienced ones. It is also observed that some advanced students prefer \textit{Beginner} groups due to lack of confidence. 
In order to successfully pass the course, all students need to pass all evaluations. Evaluations do not make any assumptions about the level and are the same for all students.
%Due to specifics of the Russian scholastic itinerary, first year students at IU can be as young as 17 years old.
%TODO (VR) report here the percentage of foreigner students
As overall structure, the course cover the foundations of imperative programming, from the notion of variable to control flow structure, but keeps tightly an object orientation introducing very early concepts as classes, objects and methods. Soon enough inheritance and polymorphism are also introduced. There is nothing new in exposing millenials to an OOP language as first programming language. These kind of experiments appeared as early as in the 80s and become very common in the 90s. The peculiarity of our approach is exposing millenials to the notion of \textit{Software Contract}  using the metaphor of business contract. Design by Contract (DbC) \cite{Meyer:1997} is an approach to achieve the so-called correctness by construction \cite{Chapman:2006}. Correctness by Construction makes use of foundations of logic, concepts that are taught by a \textit{Discrete Math} course which our students need to attend either prior or in parallel to our course. \textit{Introduction to Programming I} is not a course project (somehow along the Russian academic tradition), the evaluation is based on a set of smaller assignments, a mid-term exam and a final exam. It has to be admitted that this represents somehow a limitation, since it is difficult to relate the course with the activity of the companies already based in Innopolis.There are also quizzes that are not graded and their purpose is to provide feedback to students.
%We will come back to this point at the end of the article. 
%TODO (VR): studetns (foreigners) ~4\% (checking info right now) 

The notion of contract is introduced in the very first weeks of the course as an instrument to embed specification into the code and being sure that such a specification is checked for violation at run time. Tools for static verification also exist, for example Autoproof \cite{Tschannen:2015}, however these are only introduced towards the end of the course. The natural perception for students, at least to those who have been exposed to programming languages before, is that specification and code do not go together. It has been observed that students do not initially understand the reason of using math and logic concepts to specify the behavior of code. The perception changes when they can actually see the importance of specifying the `what' so to properly implement the `how': specification and code are not two separate artifacts, they go together and proceed together and we can automatically trace and verify their consistency. This is the very idea of formal methods, and it is something our students have been never exposed to before in the totality of cases.

%What the students perceive since the early stages is that specification and code are not two separate artifacts, they go together and proceed together and we can automatically trace and verify their consistency. This is the very idea of formal methods, and it is something our students have been never exposed to before in the totality of cases.

Students are introduced the idea of DbC without making any neat distinctions (syntactical or semantical) between the code itself and the contract, in fact a contract is presented for what it is: part of the code integrating with the imperative aspects and the modularization and reuse of code peculiar of OOP (natively supported by Eiffel). It has been noticed that students with no programming experience absorb this fact without any problem or objections, and indeed in a completely natural way. This is not true for students with a bias due to occasional and superficial previous programming experience. The bias is even stronger in students that consider themselves fluent programmers in some other language. This observation reinforces the fact that formal methods can more easily thought when there is no a priori bias. The course also introduces the concept of testing, its taxonomies and different approaches for testing measurements. Students quickly grasp the idea of using contracts as unit tests for features. It was noticed that advanced students, after having understood the main idea behind DbC, often ask about the use of contracts in the automatic generation of input values for test cases for features. They are also curious about the necessity of testing in the presence of contracts since contracts can be used for the formal verification of correctness. This observation reinforces the fact that students grasp the concepts and master them in a natural way. 

In order to understand how the course, and in particular the notion of DbC, helps students to better grasp programming concepts, a questionnaire was given to those who passed Introduction to Programming in Fall 2016. We asked a single question:

\begin{itemize}
\item Did Design by Contract help you to grasp better software concepts presented in the continuation of your study? (definitely not/ not /neutral / yes/ definitively yes)
%\item Q2: Did DbC help you to grasp better mathematical concepts presented in the continuation of your study? (definitely not/ not /neutral / yes/ definitively yes)
\end{itemize}

\begin{figure}
\includegraphics[width=3.5in]{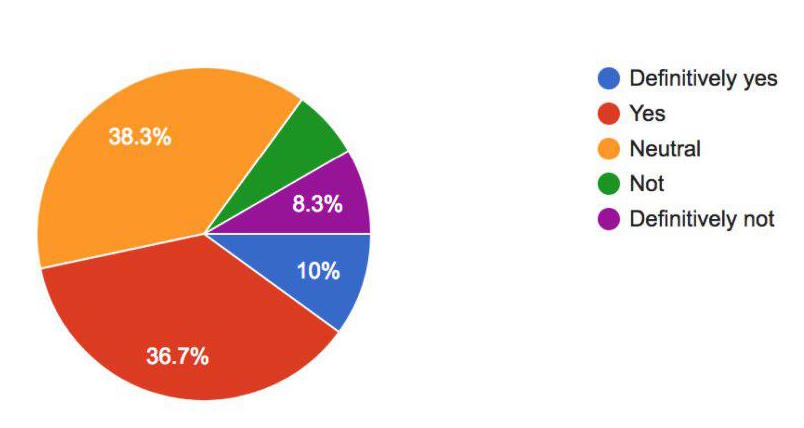}
\caption{Results of the question: Did DbC help you to better grasp software concepts presented in the continuation of your study?}
\label{fig:survey}
\end{figure}

% \begin{figure}
% \includegraphics[width=4in]{Q2}
% \caption{Results of Q2}
% \end{figure}
Fig.\ref{fig:survey} shows that students found DbC useful to better grasp programming concepts.
% while they did not consider it useful for mathematical concepts. %It has to be mentioned  that BS2 students at Innopolis have not been yet exposed to Theory of Computation, compilers or any course on software Verification.

%----------------------------------------------------
\section{Results}  \label{Results}
%----------------------------------------------------
This Section is devoted to present the data and analysis of the students' performance for the \textit{Introduction to Programming} course after implementing the choices and structure described in previous sections. We analyzed the data of students' grades who attended the course at Innopolis University and asked ourself whether the separation in beginners and advanced is useful to the pedagogical process. In this paper, we do not analyze the effect of the language division: English vs. Russian. This aspect will be explored in the future.

\begin{figure}
\includegraphics[width=3.3in]{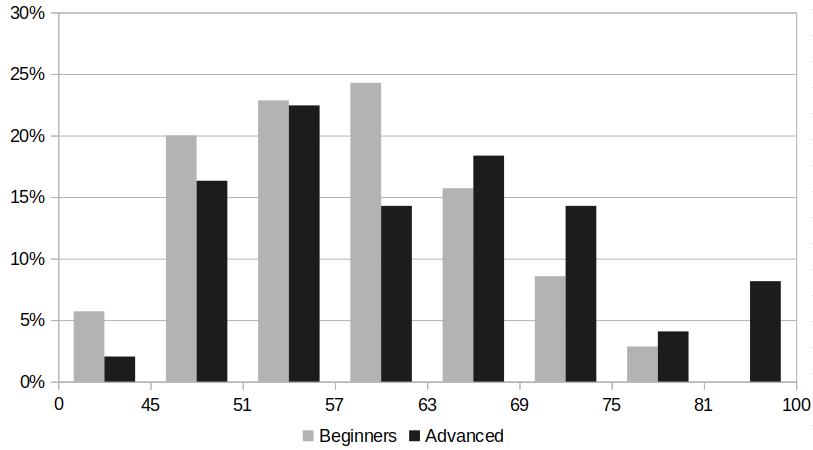}
\caption{Total grade distribution in \textit{beginners} and \textit{advanced} groups}
\label{histogram}
\end{figure}

In Fig.\ref{histogram} we report the distribution of the overall final grade of the course for Fall 2016. It is noticeable that higher grades go to advanced and lower to beginners. This suggests that the self assessment is informative for the continuation of the course and labs at different levels can be treated separately. Instructors for example can assume a better understanding of programming for advanced students. Table \ref{tab:cumulative_results} provides some further confirmation of this fact and summarizes the results of the major grading milestones for Fall 2016. This data suggests that the self assessment is sound, i.e. students are able effectively to capture their programming skills. In particular, in the final exam advanced performed about 10\% better on average. It is worth noticing however, that the best grade was obtained instead by an outlier, i.e. a beginner student who performed better than anyone else. The presence of outliers does not invalidate the general scheme. To the contrary, it is expected than some students decide to attend a group different from the self perceived level for different reasons, for example to study with a friend or to have a more comfortable environment. The presence of successful beginners can also be explained in terms of attitude. As discusses in Section \ref{DbC}, it has been observed how students that consider themselves good programmers have a bias against learning a new language and a new methodology, while beginners naturally absorb new ideas. Clearly, this bias may end up in being an inhibitor of success.

\begin{figure}
\hspace*{-1cm}
\includegraphics[width=4in]{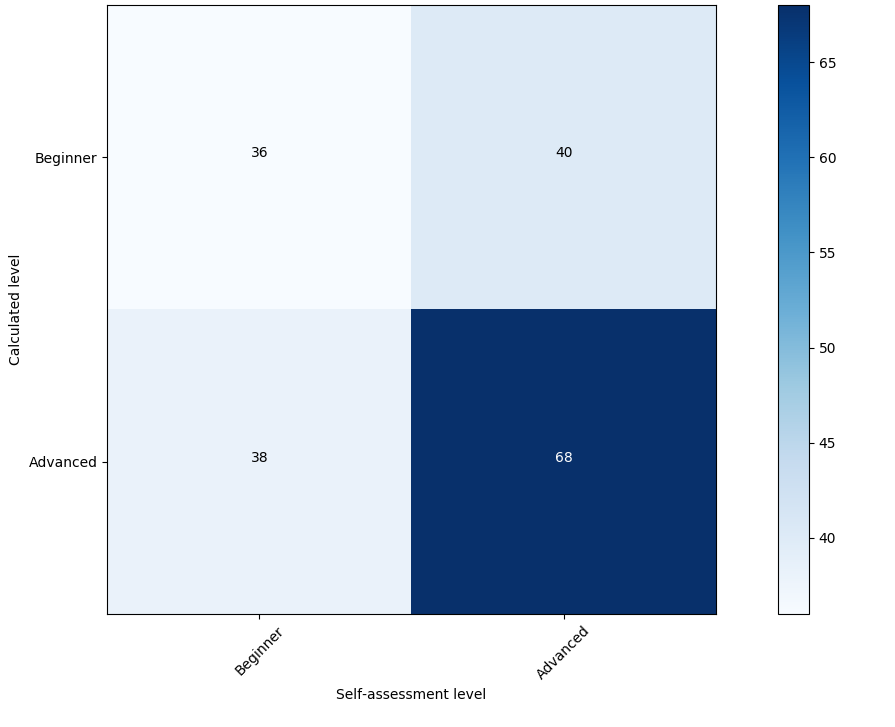}
\caption{Self-assessment vs. entry test}
\label{SAET}
\end{figure}

The data collected in Fall 2016 and reported here seemed to suggest that the division Beginner/Advanced was informative. To find more evidence of that, in Fall 2017 we compared the self assessment with the results of an actual entry test that students undertook at the beginning of the semester. The test was about computer science in general, and in particular on programming and data structures. Students were never informed about the results of such a test, so that their self assessment was not biased. Figure \ref{SAET} reports on how accurate the students' self-assessment (x-axis) compares with the actual test assessment (y-axis), for instance, 68 students who assessed themselves as \textit{advanced} were in fact advanced students based on their grades of the course. Results show with a higher level of confidence the usefulness of the division beginner/advanced.

%Grades:\\
% Advanced - 24.5\% A; 32.7\% B; 40.8\% C; 2\% D \\
% Beginner - 8.6\% A; 40\% B; 45.7\% C; 5.7 D \\

%\subsection{Did grades received over the semester support or not the validity of self assessment? - SASHA's TABLE}

%Table. \ref{cumulative} shows that, with the exception of the first assignment, advanced tend to perform better, on average. This shows that self-assessment is able to capture to a large extent the actual programming skills of students. Max and min also shows the same tendency, in particular for major milestones (mid-term and final).

% Table generated by Excel2LaTeX from sheet 'Лист1'
% Table generated by Excel2LaTeX from sheet 'Лист1'
\begin{table}[htbp]
  \centering
  \caption{Cumulative results}
  \label{cumulative}
    \begin{tabular}{|c|c|c|c|p{4.855em}|}
    \toprule
          & \textbf{Average} & \textbf{Max} & \textbf{Min} & \multicolumn{1}{c|}{} \\
    \midrule
%     \multicolumn{1}{|c|}{\multirow{3}[6]{*}{\textbf{Assignment 1}}} & 5.35  & 6     & 2.34  & \textbf{Beginners} \\
% \cmidrule{2-5}          & 5.25  & 6     & 2.94  & \textbf{Advanced} \\
% \cmidrule{2-5}          & -1.83\% & 0.00\% & 25.64\% & \textbf{Difference} \\
%     \midrule
    \multicolumn{1}{|c|}{\multirow{3}[6]{*}{\textbf{Mid-term Exam}}} & 21.11 & 32.4  & 7.6   & \textbf{Beginners} \\
\cmidrule{2-5}          & 23.80 & 37.2  & 11.2  & \textbf{Advanced} \\
\cmidrule{2-5}          & 12.70\% & 14.81\% & 47.37\% & \textbf{Difference} \\
    \midrule
%     \multicolumn{1}{|c|}{\multirow{3}[6]{*}{\textbf{Assignment 2}}} & 4.15  & 5.76  & 0     & \textbf{Beginners} \\
% \cmidrule{2-5}          & 4.45  & 5.73  & 0     & \textbf{Advanced} \\
% \cmidrule{2-5}          & 7.26\% & -0.52\% & 0     & \textbf{Difference} \\
%     \midrule
%     \multicolumn{1}{|c|}{\multirow{3}[6]{*}{\textbf{Assignment 3}}} & 5.35  & 8     & 0     & \textbf{Beginners} \\
% \cmidrule{2-5}          & 5.63  & 7.92  & 0     & \textbf{Advanced} \\
% \cmidrule{2-5}          & 5.37\% & -1.00\% & 0     & \textbf{Difference} \\
%     \midrule
    \multicolumn{1}{|c|}{\multirow{3}[6]{*}{\textbf{Final Exam}}} & 20.28 & 31.8  & 3     & \textbf{Beginners} \\
\cmidrule{2-5}          & 22.21 & 30    & 12    & \textbf{Advanced} \\
\cmidrule{2-5}          & 9.55\% & -5.66\% & 300.00\% & \textbf{Difference} \\
    \midrule
%     \multicolumn{1}{|c|}{\multirow{3}[6]{*}{\textbf{Total}}} & 56.41 & 79.13 & 16.53 & \textbf{Beginners} \\
% \cmidrule{2-5}          & 61.90 & 91.07 & 42.95 & \textbf{Advanced} \\
% \cmidrule{2-5}          & 9.73\% & 15.09\% & 159.83\% & \textbf{Difference} \\
    %\bottomrule
    \end{tabular}%
  \label{tab:cumulative_results}%
\end{table}%

\section{Discussion} \label{Discussion}
%----------------------------------------------------

\textit{Introduction to Programming} divides students in four groups on the basis of a technical self assessment and a choice of preferred teaching language (Russian or English). In this paper we analyzed only the division based on programming skills while the language choice will be investigated in future. The data collected and analyzed supports the idea that such an organization is useful, i.e. the students self assessment is a good predictor of technical skills, with the exception of some outliers that may be explained by the bias of advanced against learning a new language. The results of this paper are however not fully conclusive, and further studies are required. On top of this conclusion, we report the result of a questionnaire that shows how the majority of students who passed the course believe that DbC is useful to better grasp software concepts presented in the continuation of Innopolis curriculum. All this supports the idea that the course is effective, both in terms of content and methodological approach, and for what concerns the actual organization. After further confirmation of the results we plan to introduce changes to the course. In particular, we plan to inform students of their grades in the technical test. This will allow them to make an informed choice when they self assess their technical skills. Participation to group may remain up to students' decision, but this decision will be supported by objective information.

% \begin{itemize}
% \item Conclusion of the analysis: the division is useful, further study is required
% \item How conclusion relates to Section Intro
% \item Discussion of plans for changes in the course structure: informing students of test grades in order to do an informed self assessment
% \item Future work: study the language division
% \end{itemize}

%----------------------------------------------------
\bibliographystyle{splncs03}
\bibliography{bibl}

\begin{thebibliography}{10}
\providecommand{\url}[1]{\texttt{#1}}
\providecommand{\urlprefix}{URL }

\bibitem{java:00}
Arnold, K., Gosling, J., Holmes, D.: The Java Programming Language.
  Addison-Wesley Longman Publishing Co., Inc., Boston, MA, USA, 3rd edn. (2000)

\bibitem{Bratko:1986}
Bratko, I.: Prolog Programming for Artificial Intelligence. Addison-Wesley
  Longman Publishing Co., Inc., Boston, MA, USA (1986)

\bibitem{Chapman:2006}
Chapman, R.: Correctness by construction: A manifesto for high integrity
  software. In: Proceedings of the 10th Australian Workshop on Safety Critical
  Systems and Software - Volume 55. pp. 43--46. SCS '05, Australian Computer
  Society, Inc., Darlinghurst, Australia, Australia (2006),
  \url{http://dl.acm.org/citation.cfm?id=1151816.1151820}

\bibitem{Hejlsberg:2010}
Hejlsberg, A., Torgersen, M., Wiltamuth, S., Golde, P.: C\# Programming
  Language. Addison-Wesley Professional, 4th edn. (2010)

\bibitem{Jensen:1974}
Jensen, K., Wirth, N.: PASCAL User Manual and Report. Springer-Verlag New York,
  Inc., New York, NY, USA (1974)

\bibitem{Kernighan:1988}
Kernighan, B.W.: The C Programming Language. Prentice Hall Professional
  Technical Reference, 2nd edn. (1988)

\bibitem{Kondratyev:13}
Kondratyev, D., Tormasov, A., Stanko, T., Jones, R.C., Taran, G.: Innopolis
  university-a new it resource for russia. In: 2013 International Conference on
  Interactive Collaborative Learning (ICL). pp. 841--848 (Sept 2013)

\bibitem{Meyer:1992b}
Meyer, B.: Applying "design by contract". Computer  25(10),  40--51 (Oct 1992),
  \url{http://dx.doi.org/10.1109/2.161279}

\bibitem{Meyer:1992}
Meyer, B.: Eiffel: The Language. Prentice-Hall, Inc., Upper Saddle River, NJ,
  USA (1992)

\bibitem{Meyer:1997}
Meyer, B.: Object-oriented Software Construction (2Nd Ed.). Prentice-Hall,
  Inc., Upper Saddle River, NJ, USA (1997)

\bibitem{Milner:1997}
Milner, R., Tofte, M., Macqueen, D.: The Definition of Standard ML. MIT Press,
  Cambridge, MA, USA (1997)

\bibitem{Odersky:2004}
Odersky, M., Micheloud, S., Mihaylov, N., Schinz, M., Stenman, E., Zenger, M.,
  et~al.: An overview of the scala programming language. Tech. rep. (2004)

\bibitem{Pedroni:2006}
Pedroni, M., Meyer, B.: The inverted curriculum in practice. SIGCSE Bull.
  38(1),  481--485 (Mar 2006), \url{http://doi.acm.org/10.1145/1124706.1121493}

\bibitem{Jones:2003}
Peyton~Jones, S.: Haskell 98 language and libraries: the Revised Report.
  Cambridge University Press (2003)

\bibitem{Steele:1990}
Steele, Jr., G.L.: Common LISP: The Language (2Nd Ed.). Digital Press, Newton,
  MA, USA (1990)

\bibitem{Tschannen:2015}
Tschannen, J., Furia, C.A., Nordio, M., Polikarpova, N.: Autoproof: Auto-active
  functional verification of object-oriented programs. In: 21st International
  Conference on Tools and Algorithms for the Construction and Analysis of
  Systems. Lecture Notes in Computer Science, Springer (2015)

\bibitem{Wegner:1990}
Wegner, P.: Concepts and paradigms of object-oriented programming. SIGPLAN OOPS
  Mess.  1(1),  7--87 (Aug 1990),
  \url{http://doi.acm.org/10.1145/382192.383004}

\end{thebibliography}
\end{document}